\documentclass[aps,prl,twocolumn,groupedaddress,amsmath,amssymb]{revtex4}

\def\beq{\begin{equation}}
\def\eeq{\end{equation}}
\newcommand{\veps}{\varepsilon}
\newcommand{\christoffel}[3]{\genfrac{\{}{\}}{0pt}{}{#1\hfill}{#2 #3}}

\begin{document}

\title{GAUGE GRAVITATION THEORY IN RIEMANN-CARTAN SPACE-TIME \\ AND GRAVITATIONAL INTERACTION
}

\author{A. V. Minkevich}

\email[]{minkav@tut.by}


 \affiliation{Department of Theoretical Physics and Astrophysics, Belarussian State University,
Minsk, Belarus}
 \affiliation{ Department of Physics and Computer Methods, Warmia
and Mazury University in Olsztyn, Poland}


\begin{abstract}
The place and physical significance of gauge gravitation theory in the
Riemann-Cartan space-time (GTRC) in the framework of gauge approach to
gravitation is discussed. Isotropic cosmology built on the base of GTRC with
general expression of gravitational Lagrangian with indefinite parameters is
considered. The most important physical consequences connected with the change
of gravitational interaction, with possible existence of limiting energy
density and gravitational repulsion at extreme conditions, and also with the
vacuum repulsion effect are discussed. The solution of the problem of
cosmological singularity and the dark energy problem as result of the change of
gravitational interaction is considered.
\end{abstract}

\maketitle

\section{INTRODUCTION}

The general relativity theory (GR) is the base of modern theory of
gravitational interaction. According to GR the metric properties of physical
space-time are more complicated by taking into account the gravitational
interaction that leads to 4-dimensional pseudo-riemannian continuum. GR allows
to describe various gravitating systems and physical phenomena in astrophysics
and astronomy including the observable Universe. At the same time GR is faced
with some principal difficulties which appear at certain conditions by
description of gravitating systems.

Gravitational field describing by metric tensor of physical space-time by means
of gravitational equations by A. Einstein has the energy-momentum tensor of
physical matter as its source. In the case of usual gravitating matter with
positive values of energy density and pressure the gravitational interaction in
the frame of GR has the character of attraction which increases with energy
density together. As result this is the cause of appearing of singular states
in cosmological models of Big Bang and black holes. The presence of singular
state at the beginning of cosmological expansion in various cosmological models
with divergent energy density and singular metric leads to the problem of the
beginning of the Universe in time - the problem of cosmological singularity
(PCS). It should be noted that while the gravitational interaction in the frame
of GR can have the repulsion character in the case of gravitating matter with
negative pressure (for example, scalar fields in inflationary models), the PCS
can not be solved in GR by means of such systems: the most part of cosmological
models remain singular.

Another principal problem of GR is connected with invisible matter components -
dark energy and dark matter, the introduction of which is necessary in GR to
explain the observable cosmological and astrophysical data. Their explanation
in the frame of GR leads to conclusion that about 96\% energy in Universe is
connected with some hypothetical kinds of matter - dark energy and dark matter,
and contribution of usual baryon matter to the energy density composes only
about 4\%. The following question appears: what is the nature of dark energy
and dark matter if they exist and do they exist at all?

Many attempts were undertaken with the purpose to solve indicated problems in
the frame of GR and candidates to quantum gravitation theory - string
theory/M-theory and loop quantum gravity as well as different generalizations
of Einstein gravitation theory (see for example \cite{1,1a,1b,2} and Refs
herein). Radical ideas connected with notions of strings, extra-dimensions,
space-time quantization etc are used in these works. Different hypothetical
media and particles with unusual properties as possible candidates for dark
energy and dark matter were introduced and discussed. Note that many existent
generalizations of Einstein theory of gravitation are based on ad hoc
introducing hypothesis and do not have solid theoretical foundation.

At the same time there is the gravitation theory built in the framework of
common field-theoretical approach including the local gauge invariance
principle, which is a natural generalization of GR and which offers
opportunities to solve its principal problems as result of the change of
gravitational interaction (in comparison with GR). It is the gravitation theory
in the Riemann-Cartan continuum $U_4$ (GTRC) -- theory in 4-dimensional
physical space-time with curvature and torsion. In the frame of gauge approach
to gravitational interaction GTRC is direct and in certain sense necessary
generalization of Einstein gravitation theory.

\section{GAUGE APPROACH TO GRAVITATION THEORY AND GTRC}

The local gauge invariance principle is one of the most important physical
principles of modern theory of fundamental physical interactions. This
principle determines the profound connection between important conserving
physical quantities and fundamental (gauge) physical fields, which are carriers
of certain physical interactions and have corresponding physical quantities as
a sources. Consistent with Yang-Mills theory the procedure of introduction of
gauge fields is transparent in the case of internal symmetry groups given in
Minkowski space-time. The situation is changed by considering the gravitational
interaction, in this case the gauge group is connected with coordinates
transformations and by their localization the geometrical structure of physical
space-time is changed. If the energy-momentum tensor is considered as source of
gravitational field, the gravitational interaction has to be introduced on the
base of localization of 4-parametric translations group in Minkowski
space-time, invariance with respect to this group leads according to Noether
theorem to energy-momentum tensor and conservation laws of energy and momentum.
The gravitational field as symmetric tensor field of the second rank was
introduced for the first time in \cite{3} exactly by this way. The introducing
gauge field was connected with metric tensor of physical space-time, which
assumed the structure of pseudo-riemannian continuum. The gravitational field
as generalized gauge field in the form of symmetric tensor field connected with
4-parametric translations group was considered also in \cite{4}. Thereby the
localization of 4-parametric translations group leads to metric gravitation
theory which is covariant with respect to general coordinates transformations
and by corresponding choice of gravitational Lagrangian comes to Einstein
gravitation theory. In \cite{5} gravitational field was introduced also by
localization of translations group, and the gauge field was presented as 4
fields connected with orthonormalized tetrad; corresponding theory is
gravitation theory in teleparallelism space-time.

Let us consider the question about the role of the Lorentz group in gravitation
theory introduced on the base of localization of 4-parametric translations
group. We are talking about the group of tetrad Lorentz transformations
appearing by the presence of orthonormalized tetrad at any spacetime point and
which is not connected with holonomic coordinate transformations. Because the
metric tensor $g_{\mu\nu}$  connected with tetrad $h^i{}_\mu$ according to
$g_{\mu\nu}=\eta_{ik} h^i{}_{\mu} h^i{}_{\nu}$ ($\eta_{ik}=diag(1,-1,-1,-1)$ is
metric tensor of Minkowski space-time, holonomic and anholonomic space-time
coordinates are denoted by means of greek and latin indices respectively) is
invariant with respect to tetrad Lorentz transformations with arbitrary
parameters, tetrad formulation of metric gravitation theory which we obtain by
introduction of orthonormalized tetrad at every space-time point is invariant
with respect to localized Lorentz group. This means that the group of tetrad
Lorentz transformations does not play the dynamical role from the point of view
of gauge approach. The disappearance of Noether invariant corresponding to the
Lorentz group in metric gravitation theory is connected with this fact
\cite{6}. In regard to gravitation theory in teleparallelism space-time this
theory is covariant with respect to tetrad Lorentz transformations with
constant parameters and corresponds to intermediate stage of construction of
theory, which is covariant with respect to localized Lorentz group. The
transition to this theory is obtained by virtue of introduction of gauge field
which has transformation properties of anholonomic Lorentz connection \cite{7}.
The interpretation of this field as independent dynamical field leads to GTRC
which is known in literature as Poincar\'e gauge theory of gravity
\footnote{Strictly speaking the gauge group of GTRC is direct product of
localized 4-parametric translations group and group of tetrad Lorentz
transformations. Note that localized 4-parametric translations group includes
arbitrary holonomic coordinates transformations inserting inhomogeneous Lorentz
transformations}.

It should be noted that at the first time the treatment of gravitational
interaction on the base of the gauge invariance principle was undertaken by R.
Utiyama in 1956 shortly after construction of Yang-Mills theory \cite{7}.
Utiyama considered the Lorentz group as gauge group, and because the
transformation properties of anholonomic Lorentz connection are the same in
riemannian and Riemann-Cartan space-time, Utiyama obtained Einstein
gravitational equations by identifying the Lorentz gauge field with Ricci
rotation coefficients of riemannian space-time. However, similar identification
is impossible, if the Lorentz gauge field is considered as independent
dynamical field \cite{8,9}. In addition, the treatment of gravitational field
as Lorentz gauge field is not consistent, if we take into account the
correspondence between gauge fields and their sources.

The principal significance of GTRC in the framework of gauge approach in theory
of gravitational interaction is determined by the role, which the Lorentz group
plays in modern physics. The invariance of physical theory with respect to
tetrad Lorentz transformations means that locally metrical physical space-time
properties coincide with that of Minkowski space-time. Besides metric
properties the physical space-time possesses properties connected with torsion
of Lorentz connection which plays the role of fundamental physical field.
Together with tetrad $h^i{}_\mu$ anholonomic Lorentz connection $A^{ik}{}_\mu =
- A^{ki}{}_\mu$ are independent gravitational field variables. Corresponding
field strengths are the torsion tensor $S^i{}_{\mu\nu}$ and the curvature
tensor $F^{ik}{}_{\mu\nu}$. Being strength corresponding to the group of tetrad
Lorentz transformations the curvature tensor is defined by the way as
Yang-Mills field strength. Unlike curvature, the torsion tensor as strength
corresponding to subgroup of space-time translations is the function not only
of tetrad and their derivatives, but also of Lorentz gauge field that is
distinguishing feature of gauge theory connected with coordinate
transformations. Gravitational Lagrangian is invariant built with the help of
the curvature and torsion tensors (by using tetrad or metric). In the case of
minimal coupling of matter with gravitational field defined by means of
replacement in matter Lagrangian (written in orthogonal cartesian coordinate
system in Minkowski space-time) of space-time metric and particular derivatives
of matter variables by covariant derivatives defined by total Riemann-Cartan
connection the role of sources of gravitational field in equations of PGTG play
the energy-momentum and spin momentum tensors of gravitating matter
\footnote{The minimal coupling of gravitational field with matter is used
except when this coupling leads to unacceptable consequences. So in the case of
definition of interaction of gravitational field with electromagnetic and
Yang-Mills fields we have to use coupling used in GR, because the minimal
coupling leads to violation of gauge invariance for these fields.}. The
simplest GTRC is Einstein-Cartan theory which corresponds to the choice of
gravitational Lagrangian in the form of scalar curvature \cite{10}.
Gravitational equations of this theory are identical to Einstein gravitational
equations of GR in the case of spinless matter, and in the case of spinning
sources Einstein-Cartan theory leads to linear relation between space-time
torsion and spin momentum of gravitating matter. Because of the fact that in
the frame of Einstein-Cartan theory the torsion vanishes in absence of spin,
the opinion that the torsion is generated only by spin momentum of gravitating
matter is widely held in literature. However, such situation seems unnatural,
if we take into account that the torsion tensor plays the role of gravitational
field strength corresponding to subgroup of space-time translations connected
directly in the frame of Noether formalism with energy-momentum tensor and,
consequently, the torsion can be created by spinless matter. The situation
comes to normal by including to gravitational Lagrangian similarly to theory of
Yang-Mills fields terms quadratic in gauge gravitational field strengths - the
curvature and torsion tensors, and GTRC is gravitation theory, in the frame of
which the gravitational field is described by means of interacting metric and
torsion fields and created by energy-momentum and spin momentum of gravitating
matter (see \cite{8,9,10,11,12,13,14,15,16,17}).

There are various generalizations of GTRC connected with using other groups
instead of the Lorentz group -- conformal gauge theory, (anti-) de Sitter gauge
theory, affine-metric gauge theory, in the frame of which connection possesses
in addition to torsion also nonmetricity. In comparison with similar
generalizations the principal importance of GTRC is determined by fundamental
role of the Lorentz group in physics and first of all in theory of fundamental
physical interactions.

\section{GRAVITATION EQUATIONS OF GTRC, CORRESPONDENCE PRINCIPLE OF PGTG WITH EINSTEIN GRAVITATION THEORY}

As it was noted above in the framework of GTRC the role of gravitational field
variables play the orthonormalized tetrad $h^i{}_\mu$ and the Lorentz
connection $A^{ik}{}_\mu$; corresponding field strengths are the torsion tensor
$S^i{}_{\mu\nu}$ and the curvature tensor $F^{ik}{}_{\mu\nu}$ defined as
\[
S^i{}_{\mu \,\nu }  = \partial _{[\nu } \,h^i{}_{\mu ]}  - h_{k[\mu }
A^{ik}{}_{\nu ]}\,,
\]
\[
F^{ik}{}_{\mu\nu }  = 2\partial _{[\mu } A^{ik}{}_{\nu ]}  + 2A^{il}{}_{[\mu }
A^k{}_{|l\,|\nu ]}.\
\]
The structure of gravitational equations of GTRC depends on the choice of
gravitational Lagrangian $\mathcal{L}_{\rm g}$. Because quadratic part of
gravitational Lagrangian is unknown, we will consider the GTRC based on
gravitational Lagrangian given in the following sufficiently general form
corresponding to spacial parity conservation
\begin{eqnarray}\label{1}
\mathcal{L}_{\rm g}=  f_0\,
F+F^{\alpha\beta\mu\nu}\left(f_1\:F_{\alpha\beta\mu\nu}+f_2\:
F_{\alpha\mu\beta\nu}+f_3\:F_{\mu\nu\alpha\beta}\right)  \nonumber \\
+ F^{\mu\nu}\left(f_4\:F_{\mu\nu}+f_5\: F_{\nu\mu}\right) +
f_6\:F^2 \nonumber \\
+S^{\alpha\mu\nu}\left(a_1\:S_{\alpha\mu\nu}+a_2\: S_{\nu\mu\alpha}\right)
+a_3\:S^\alpha{}_{\mu\alpha}S_\beta{}^{\mu\beta},
\end{eqnarray}
where $F_{\mu\nu}=F^{\alpha}{}_{\mu\alpha\nu}$, $F=F^\mu{}_\mu$, $f_0=(16\pi
G)^{-1}$, $G$ is Newton's gravitational constant (the light speed in the vacuum
$c=1$), $f_i$ ($i=1,2,\ldots,6$), $a_k$ ($k=1,2,3$) are indefinite parameters
\footnote{It should be noted that one out of three parameters $f_3$, $f_5$ and
$f_6$ can be excluded because of relation $\delta\int
[F_{\mu\nu\alpha\beta}F^{\alpha\beta\mu\nu}-4F_{\nu\mu}F^{\mu\nu}+F^2]h d^4
x=0$.}. Gravitational equations of GTRC obtained from the action integral $ I =
\int {\left( {{\cal L}_g + {\cal L}_m } \right)\,}h d^4 x$, where
$h=\det{\left(h^i{}_\mu\right)}$ and ${\cal L}_m$ is the Lagrangian of
gravitating matter, contain the system of 16+24 equations corresponding to
gravitational variables $h^i{}_\mu$ and $A^{ik}{}_\mu$:
\begin{eqnarray}\label{2}
\nabla_{\nu}U_{i}{}^{\mu\nu}+2S^k{}_{i\nu}U_k{}^{\mu\nu}+2(f_0+2f_6\:F)F^{\mu}{}_i   \nonumber \\
+4f_1\:F_{klim}F^{kl\mu m}+4f_2\:F^{k[m\mu]l}F_{klim}   \nonumber \\
+4f_3\:F^{\mu klm}F_{lmik}+2f_4(F_{ki}F^{k\mu}+F^{\mu}{}_{kim}F^{km})  \nonumber \\
+2f_5(F_{ki}F^{\mu k}+F^{\mu}{}_{kim}F^{mk})-h_{i}{}^{\mu}{\cal
L}_g=-T_{i}{}^{\mu},
\end{eqnarray}
\begin{eqnarray}\label{3}
4\nabla_{\nu}[(f_0/2+f_6\:F)h_{[i}{}^{\nu}h_{k]}{}^{\mu}+f_1\:F_{ik}{}^{\nu\mu}   \nonumber \\
+f_2\:F_{[i}{}^{[\nu}{}_{k]}{}^{\mu]}+f_3\:F^{\nu\mu}{}_{ik}+f_4\:F_{[k}{}^{[\mu}h_{i]}{}^{\nu]}+  \nonumber \\
+f_5\:F^{[\mu}{}_{[k}h_{i]}{}^{\nu]}]+U_{[ik]}{}^{\mu}=-J_{[ik]}{}^{\mu},
\end{eqnarray}
where
$U_{i}{}^{\mu\nu}=2(a_1\:S_{i}{}^{\mu\nu}-a_2\:S^{[\mu\nu]}{}_{i}-a_3\:S_{\alpha}{}^{\alpha
[\mu }h_{i}{}^{\nu]})$,  $T_{i}{}^{\mu}=-\frac {1}{h} \frac{\delta({h \cal
L}_m)} {\delta h^{i}{}_{\mu}}$, $J_{[ik]}{}^{\mu}=-\frac {1}{h} \frac{\delta({h
\cal L}_m)} {\delta A^{ik}{}_{\mu}}$, $\nabla_{\nu}$ denotes the covariant
operator having the structure of the covariant derivative defined in the case
of tensor holonomic indices by means of Christoffel coefficients
$\christoffel{\lambda}{\mu}{\nu}$ and in the case of tetrad tensor indices by
means of anholonomic Lorentz connection $A^{ik}{}_{\nu}$ (for example
$\nabla_{\nu} h^{i}{}_{\mu}=\partial _{\nu } \,h^i{}_{\mu
}-\christoffel{\lambda}{\mu}{\nu}\, h^{i}{}_{\lambda}-A^{ik}{}_{\nu}h_{k\mu}$).
By using minimal coupling of gravitational field with matter the tensors
$T_{i}{}^{\mu}$ and $J_{[ik]}{}^{\mu}$ are the energy-momentum and spin
momentum tensors of gravitating matter. Gravitational equations (2)-(3) are
complicated system of differential equations in partial derivatives with
indefinite parameters $f_i$ and $a_k$. Physical consequences depend essentially
on restrictions on these parameters. Some of such restrictions were obtained by
investigation of isotropic cosmology built in the frame of GTRC with
gravitational Lagrangian (1) (see below).

In order to establish the fulfilment of correspondence principle of GTRC with
Einstein gravitation theory, gravitational equations (2)-(3) will be considered
in linear approximation. In accordance with \cite{16} equations (2) in linear
approximation in metric and torsion by taking into account (3) do not contain
higher derivatives of metric functions if the following restrictions are
fulfilled
\begin{eqnarray}\label{4}
a = 2a_1  + a_2  + 3a_3=0,
\nonumber\\
4(f_1+\frac{f_2}{2}+f_3)+f_4+f_5=0.
\end{eqnarray}
Then equations for the functions $h_{\mu\nu}$
($g_{\mu\nu}=\eta_{\mu\nu}+h_{\mu\nu}$) take the form
\begin{eqnarray}\label{5}
G_{\mu\nu}^{(1)}=\frac{1}{2f_0} T_{\mu\nu}^{sym}+ \alpha (\eta_{\mu\nu} \square
-\partial_{\mu} \partial_{\nu})T,
\end{eqnarray}
where $G_{\mu\nu}^{(1)}$ is Einstein tensor in linear approximation with
respect to $h_{\mu\nu}$, $T_{\mu\nu}$ is canonical energy-momentum tensor in
Minkowski spacetime, $T=\eta^{\mu\nu} T_{\mu\nu}$, $T_{\mu\nu}^{sym}$ is
symmetrized energy-momentum tensor, $\square$ is d'Alembert operator and
parameter $\alpha=\frac{f}{3f_0^2}$, where $f = f_1  + \frac{{f_2 }} {2} + f_3
+ f_4 + f_5 + 3f_{6}>0$, has inverse dimension of energy density. According to
(5) equations of GTRC in linear approximation lead to Einstein equations for
the metric if $\alpha T \ll1$. This condition restricts acceptable energy
densities if the value $\alpha^{-1}$ corresponds to extremely high energy
densities. Exactly such situation takes place in the frame of isotropic
cosmology, where the parameter $\alpha^{-1}$ determines the value of limiting
energy density (see below). As result the correspondence of GTRC to Einstein
gravitation theory takes place in linear approximation excepting gravitating
systems with extremely high energy densities (for example massive stars
collapsing in the frame of GR). It should be noted that correspondence GTRC to
GR can take place if the second condition (4) for parameters $f_{i}$ is not
valid. Then the equations (5) acquire additional terms with higher derivatives
of $h_{\mu\nu}$ that leads to appearance in expression of gravitational
potential $\phi$ for material point of mass $M$ additional Yukawa-type term
\cite{16}
\begin{equation}\label{5a}
 \phi =- \frac{GM}{r}[1 + k exp(-mr)],
\end{equation}
where constants $k$ and $m$ are some functions of indefinite parameters $f_{i}$
and $a_{k}$. In the case if $0<k \ll1$ physical consequences of GTRC and GR
practically coincide.

While GTRC corresponds to GR in linear approximation, conclusions of GTRC and
GR in non-linear regime at cosmological and astrophysical scales can be
essentially different. Similar differences are demonstrated below in the case
of isotropic cosmology built in the frame of GTRC.

\section{GTRC AND ISOTROPIC COSMOLOGY, COSMOLOGICAL EQUATIONS AND EQUATIONS FOR TORSION FUNCTIONS}

The structure of gravitational equations of GTRC (2)-(3) is simplified in the
case of gravitating systems with high spacial symmetry, then the number of
gravitational equations and their dependence on indefinite parameters are
reduced. The symmetry of homogeneous isotropic models (HIM) which are used in
the frame of isotropic cosmology is given by set of six Killing vectors (see
for example \cite{20}). According to Killing equations the space-time metric is
given by Robertson-Walker metric which by choosing spherical coordinate system
is: $g_{\mu \,\nu }= diag (1, -\frac {R^2}{1-k r^2}, -R^2 r^2, -R^2 r^2
sin^2{\theta})$, where  $R(t)$ is the scale factor of Robertson-Walker metric
and $k=0, +1, -1$ for flat, closed and open models respectively. The structure
of torsion tensor determined from condition of vanishing of Lie derivatives
relative to Killing vectors is given by two torsion functions $S_{1}(t)$ and
$S_{2}(t)$ determining the following non-vanishing components of torsion tensor
(with holonomic indices) \cite{21, 22}: \begin{eqnarray}\label{7}
S^1{}_{10}=S^2{}_{20}=S^3{}_{30}=S_{1}(t),
\nonumber\\
S_{123}=S_{231}=S_{312}=S_{2}(t)\frac{R^3r^2}{\sqrt{1-kr^2}}\sin{\theta}.
\end{eqnarray}
By choosing the tetrad corresponding to Robertson-Walker metric (6) in diagonal
form and by using (7) we find the Lorentz connection and following
non-vanishing tetrad components of curvature tensor noted by sign \^{} :
\begin{eqnarray}
&&
  F^{\hat 0\hat 1}{}_{\hat 0\hat 1}  = \,F^{\hat 0\hat 2}{}_{\hat 0\hat 2}
    = F^{\hat 0\hat 3}{}_{\hat 0\hat 3}
    \equiv A_1,
  F^{\hat 1\hat 2}{}_{\hat 1\hat 2}
    = F^{\hat 1\hat 3}{}_{\hat 1\hat 3}  = F^{\hat 2\hat 3}{}_{\hat 2\hat 3}
    \equiv A_2, 
\nonumber\\
&&
  F^{\hat 0\hat 1}{}_{\hat 2\hat 3}  = \,F^{\hat 0\hat 2}{}_{\hat 3\hat 1}
    = F^{\hat 0\hat 3}{}_{\hat 1\hat 2}
    \equiv A_3,
  F^{\hat 3\hat 2}{}_{\hat 0\hat 1} = F^{\hat 1\hat 3}{}_{\hat 0\hat 2}
    = F^{\hat 2\hat 1}{}_{\hat 0\hat 3} \equiv A_4, 
\nonumber 
\end{eqnarray}
\begin{equation}\label{2}
\begin{aligned}
&
    A_1=\dot{H}+H^2-2HS_1-2\dot{S}_1,
\\
&    A_{2}  = \frac{k} {{R^2 }} + \left( {H - 2S_1 } \right)^2  - S_2^2,
\\
&    A_{3}  = 2\left( {H - 2S_1 } \right)S_2,
\\
&    A_{4}  = \dot S_2+HS_2,
\end{aligned}
\end{equation}
where $H=\dot{R}/R $ is Hubble parameter and a dot denotes the differentiation
with respect to time. Bianchi identities for 4-dimensional Riemann-Cartan
space-time
\begin{equation}
\varepsilon^{\sigma\lambda\mu\nu}\nabla_\lambda
 F^{ik}{}_{\mu\nu}=0
\end{equation}
are reduced in the case of HIM to two following relations \cite{23}:
\begin{eqnarray}\label{7}
    \dot A_{2}  + 2H\left( {A_{2}  - A_{1} } \right) + 4S_1 A_{1}
        + 2S_2 A_{4}  = 0, \nonumber\\
\dot A_{3}  + 2H\left( {A_{3}  - A_{4} } \right) + 4S_1 A_{4}
        - 2S_2 A_{1}  = 0.
\end{eqnarray}
The system of gravitational equations (2)-(3) in the case of HIM is reduced to
the system of 4 differential equations, which can be written as \cite{23}:
\begin{eqnarray}\label{5}
a\left( {H - S_1 } \right)S_1  - 2bS_2^2  - 2f_0 A_{2}  + 4f\left( {A_{1}^2 -
A_{2}^2 } \right) \nonumber \\ + 2q_2 \left( {A_{3}^2 - A_{4}^2 } \right) =  -
\frac{\rho } {3},
\end{eqnarray}
\begin{eqnarray}\label{6}
a\left( {\dot S_1  + 2HS_1  - S_1^2 } \right) - 2bS_2^2  - 2f_0 \left( {2A_{1}
+ A_{2} } \right) \nonumber \\ - 4f\left( {A_{1}^2 - A_{2}^2 } \right) - 2q_2
\left( {A_{3}^2  - A_{4}^2 } \right) = p,
\end{eqnarray}
\begin{eqnarray}\label{7}
f\left[ {\dot A_{1}  + 2H\left( {A_{1}  - A_{2} } \right) + 4S_1 A_{2} }
\right] + q_2 S_2 A_{3} \nonumber \\ - q_1 S_2 A_{4}  + \left( {f_0  + \frac{a}
{8}} \right)S_1  = 0,
\end{eqnarray}
\begin{eqnarray}\label{8}
q_2 \left[ {\dot A_{4}  + 2H\left( {A_{4}  - A_{3} } \right) + 4S_1 A_{3} } \right] - \Big [ 4f\, A_{2} \nonumber \\
+ 2q_1  A_{1}  + \left( {f_0  - b} \right) \Big] S_2  = 0.
\end{eqnarray}
where
\begin{eqnarray}
  a = 2a_1  + a_2  + 3a_3, \qquad b = a_2  - a_1,
\nonumber\\
  f = f_1  + \frac{{f_2 }} {2} + f_3  + f_4  + f_5  + 3f_{6}\, ,
\nonumber\\
  q_1  = f_2  - 2f_3  + f_4  + f_5  + 6f_{6}, \qquad q_2  = 2f_1  - f_2 ,
\nonumber
\end{eqnarray}
$\rho$ and $p$ are the energy density and the pressure of gravitating matter
respectively, and average value of spin momentum is supposed to be equal to
zero.

The system of gravitational equations for HIM (11)-(14) contains in general
case 5 indefinite parameters and allows to obtain cosmological equations and
equations for torsion functions. Without using any restrictions on indefinite
parameters we obtain the following expressions for curvature functions $A_1$
and $A_2$ \cite{24}:
\begin{eqnarray}\label{12}
A_1=-\frac{1} {{12(f_0 +a/8)
Z}}
        \Big[
            \rho  + 3p - \frac{2f}{3} F^2 + 8 q_2 FS_2^2
\nonumber\\
      - 12q_2 \left( {\left( {HS_2  + \dot S_2 } \right)^2
                + 4\left( {\frac{k}{{R^2 }} - S_2^2 } \right)S_2^2 }
            \right) \nonumber\\ - \frac{3a} {2}
            \left( \dot{H} + H^2 \right)
            \Big],
\nonumber\\
A_2=\frac{1} {{6(f_0 +a/8)Z}}
       \Big[
            \rho  - 6 (b +a/8)S_2^2 + \frac{f}{3} F^2
\nonumber\\
        + \frac{3a} {4}
            \left({\frac{k}{{R^2 }} + H^2}\right) \nonumber\\
             - 6 q_2 \left({\left( {HS_2  + \dot S_2 } \right)^2
                + 4\left( {\frac{k}{{R^2 }} - S_2^2 } \right)S_2^2
                }\right)
           \Big],
\end{eqnarray}
where scalar curvature
 \begin{eqnarray}\label{11}
         F=\frac{1}{2(f_0 + a/8)} \Big[
                \rho-3p - 12(b+a/8) S_2^2
\nonumber \\
                + \frac{3a}{2} \left(\frac{k}{R^2}+\dot{H}+2H^2\right)
            \Big]
\end{eqnarray}
and $Z=1 + \frac{1} {(f_0+ a/8)} \left(\frac{2f} {3} F - 4q_2 S_2^2\right)$. We
obtain the generalization of Friedmann cosmological equations by substituting
in definitions (8) of functions $A_1$ and $A_2$ their expressions (15) found
from gravitational equations for HIM. These equations contain the torsion
functions $S_1$ and $S_2$, which can be found from gravitational equations by
using Bianchi identity (10) and definition of the curvature functions $A_3$ and
$A_4$. As result the torsion function $S_1$ takes the following form:
\begin{eqnarray} \label{17}
    S_1  = -\frac{1}{6 (f_0 + a/8)Z} 
    [f \dot F  \nonumber \\
     + 6(2f-q_1+2q_2) H S_2^2
            +6(2f-q_1) S_2 \dot S_2],
\end{eqnarray}
and the torsion function $S_2$ satisfies the differential equation of the
second order:
\begin{eqnarray}\label{18}
    q_2 \left[ \ddot S_2  + 3H\dot S_2  + \left(3\dot{H} - 4 \dot S_1
    +4S_1(3H
        - 4 S_1)\right) S_2  \right]
\nonumber \\
        - \left[\frac{q_1+q_2} {3} F + (f_0-b)
-2 (q_1+q_2-2f) A_2 \right]S_2 & = & 0.
\end{eqnarray}
From formulas (16) and (17) for scalar curvature $F$ and torsion function $S_1$
we see that cosmological equations do not contain higher derivatives of the
scale factor $R$ if $a=0$. With the purpose to exclude higher derivatives of
$R$ from cosmological equations the restriction $a=0$ was used in our studies.
By using this restriction we will write principal relations of isotropic
cosmology of GTRC based on general expression $\mathcal{L}_{\rm g}$ (1).

Cosmological equations take the following form:
\begin{eqnarray}\label{15}
    \frac{k}{R^2} + (H-2S_1)^2= \nonumber\\
    \frac{1}{{6f_0 Z}}
        \left[
            {\rho  +6\left(f_0 Z- b\right) S_2^2
            + \frac{\alpha }{4} \left( {\rho  - 3p - 12bS_2^2 } \right)^2 }
        \right]
\nonumber\\
        - \frac{{3\alpha \veps f_0 }} {Z}
            \left[
                {\left( {HS_2  + \dot S_2 } \right)^2
                + 4\left( {\frac{k}{{R^2 }} - S_2^2 } \right)S_2^2 }
            \right],
\end{eqnarray}
\begin{eqnarray}\label{16}
    \dot{H}+H^2-2HS_1-2\dot{S}_1 = \nonumber\\
    -\frac{1} {{12f_0 Z}}
        \left[
            \rho  + 3p - \frac{\alpha } {2} \left( {\rho  - 3p - 12bS_2^2 } \right)^2
        \right]
\nonumber\\
        - \frac{\alpha \veps }{Z}\left( {\rho  - 3p - 12bS_2^2 } \right)S_2^2
\nonumber\\
        + \frac{{3\alpha \veps f_0 }} {Z}
            \left[ {\left( {HS_2  + \dot S_2 } \right)^2
                + 4\left( {\frac{k}{{R^2 }} - S_2^2 } \right)S_2^2 }
            \right],
\end{eqnarray}
where scalar curvature is $F=\frac{1}{2f_0}(\rho-3p - 12b S_2^2)$,
$Z=1+\alpha\left( \rho - 3p - 12\left( {b + \veps f_0 } \right)S_2^2\right)$,
and besides parameters $\alpha=\frac {f} {3f_0^2}$ and $b$ equations (19)-(20)
contain dimensionless parameter $\veps=q_2/f$. In accordance with (17)-(18) the
torsion functions are determined by the following way:
\begin{eqnarray}\label{17}
    S_1  = -\frac{\alpha }{4Z} [\dot \rho
    - 3 \dot p + 12f_0(3 \veps + \omega) H S_2^2
    \nonumber\\
    -12( {2b - (\veps + \omega)f_0 } ) S_2 \dot S_2],
\end{eqnarray}
\begin{eqnarray}\label{18}
    \varepsilon [ \ddot S_2  + 3H\dot S_2  + \left(3\dot{H} - 4 \dot
    S_1
    + 12 HS_1 - 16 S_1^2\right) S_2 ]  \nonumber\\
        - \frac{1} {{3f_0 }} [ ( 1- \frac {1} {2} \omega) ({\rho  - 3p - 12bS_2^2
        })  \nonumber\\
        + \frac{{\left( {1  - b/f_0}\right)}} {\alpha} + 6f_0 \omega A_2]S_2  =
        0,
\end{eqnarray}
where dimensionless
parameter $\omega= \frac {2f - q_1 - q_2} {f}$ is introduced.

Cosmological equations (19)-(20) together with equations (21)-(22) determine
the evolution of HIM in Riemann-Cartan space-time if equation of state of
gravitating matter is known. It is necessary to keep in mind that matter
content and its equation of state change by evolution of Universe, and in the
case of spinless matter minimally coupled with gravitation the equation of the
energy conservation takes the same form as in GR
\begin{equation}\label{24}
\dot{\rho}+3H\left(\rho+ p\right)=0.
\end{equation}
Obtained equations of isotropic cosmology (19)-(22) contain 4 indefinite parameters:
$\alpha$ (or $f$), $b$, $\varepsilon$ and $\omega$. These parameters have certain values by
supposing that GTRC is correct gravitation theory. We can find restrictions on indefinite
parameters by analyzing physical consequences of isotropic cosmology in dependence on
these parameters, by which these consequences are the most satisfactory and correspond to
observable cosmological data.

\section{VACUUM GRAVITATIONAL REPULSION EFFECT AND DARK ENERGY PROBLEM}

At first we will consider the behaviour of cosmological solutions at
asymptotics, where energy density is sufficiently small: $0<\omega \alpha
(\rho+3p)\ll1$. It is easy to show that the cosmological equations at
asymptotics  by certain restrictions on indefinite parameters take the form of
Friedmann cosmological equations of GR with effective cosmological constant
induced by torsion function $S_2$. This situation takes place if parameter
$\varepsilon$ is sufficiently small ($|\varepsilon|\ll 1$) and at least one of
two following conditions is valid: $|\omega|\ll 1$ or $0<1-\frac{b}{f_0}\ll 1$
together with $0<\omega<4\frac{b}{f_0}$ \cite{25}. Then the torsion function
$S_2$ according to (22) takes the form
\begin{equation}\label{24}
S_2^2  = \frac {1} {12b} \left[\rho  - 3p + \frac {1  - b/f_0} {\alpha}\right],
\end{equation}
and cosmological equations are transformed by the following way:
\begin{equation}\label{25}
    \frac{k}{R^2 } + H^2  = \frac{1}{6f_0 }\left[\rho \frac{f_0}{b} + \frac{1}{4\alpha} \left(1 - \frac{b}{f_0}\right)^2
    \frac{f_0}{b} \right],
\end{equation}
\begin{equation}\label{26}
    \dot H + H^2  =  - \frac{1} {{12f_0 }}\left[ (\rho + 3p) \frac{f_0}{b} - \frac{1}{2\alpha}
    \left(1 - \frac{b}{f_0}\right)^2 \frac{f_0}{b}\right].
\end{equation}
If the parameter $\alpha$ corresponds to the scale of extremely high energy
densities, the parameter $b$ has to satisfy the condition $0<1-\frac{b}{f_0}\ll
1$. By certain relation between parameters $\alpha$ and $b$ the effective
cosmological constant in equations (25)-(26) coincides with cosmological
constant accepted by observational data concerning acceleration of cosmological
expansion at present epoch. The appearance of effective cosmological constant
in cosmological equations at asymptotics allows to explain accelerating
cosmological expansion at present epoch without using the notion of dark energy
as result of the change of gravitational interaction provoked by space-time
torsion. It is connected with the fact that the physical space-time in the
vacuum has the structure of Riemann-Cartan continuum with de Sitter metric and
non-vanishing torsion (without introducing cosmological constant) that
demonstrates the dynamical role of the physical vacuum in the frame of GTRC
\cite{24} \footnote{It should be noted that properties of space and time are
connected with investigation of spatial and temporal relations for material
objects and processes. The geometrical space-time structure in the vacuum is
found by analyzing of HIM when energy density and pressure of matter tend to
zero \cite{24}.}. The effect of vacuum gravitational repulsion in the frame of
GTRC leading to accelerating expansion at present epoch has non-linear
character and it is essential at cosmological scale. As it was shown in
\cite{26}, cosmological solutions at asymptotics are stable if $\varepsilon>0$
.
\section{LIMITING ENERGY DENSITY AND PROBLEM OF COSMOLOGICAL SINGULARITY}

By certain restrictions on indefinite parameters cosmological equations for HIM
filled with usual gravitating matter with positive values of energy density and
pressure lead to existence of limiting (maximum) energy density, near to which
the gravitational interaction is repulsive that ensures the regularization of
cosmological solutions of such models in the frame of GTRC. At the first time
the conclusion about possible existence of limiting energy density was obtained
in the case of HIM with the only torsion function $S_1$ $(S_2=0)$ \cite{18}
(see also \cite{19}). Cosmological equations for such HIM are very simple and
depend on just one indefinite parameter $\alpha$. However, HIM with the only
torsion function possess principal drawback because of divergence of torsion at
the state with limiting energy density, but consistent description in the frame
of classical theory assumes regular behaviour of all physical quantities
including the torsion and curvature functions. In addition, it is impossible to
solve the problem of dark energy by considering these models, because the
physical space-time in the vacuum in this case has the structure of Minkowski
space-time \cite{24}. Simultaneous solution of PCS and dark energy problem in
the frame of isotropic cosmology can be obtained in the case of HIM with two
torsion functions.

The existence of limiting energy density follows strictly from eqs. (19)-(22),
if we put that the small parameter $\varepsilon$ just vanishes $\varepsilon=0$
\cite{27}, that leads to their essential simplification. Then cosmological
equations (19)-(20) take the form
\begin{eqnarray}\label{24}
    \frac{k}{R^2} + (H-2S_1)^2 -S_2^2= \nonumber\\
    \frac{1}{{6f_0 Z}}
        \left[
            {\rho  -6 b S_2^2
            + \frac{\alpha }{4} \left( {\rho  - 3p - 12bS_2^2 } \right)^2 }
        \right],
\end{eqnarray}
\begin{eqnarray}\label{25}
    \dot{H}+H^2-2HS_1-2\dot{S}_1 = \nonumber\\
    -\frac{1} {{12f_0 Z}}
        \left[
            \rho  + 3p - \frac{\alpha } {2} \left( {\rho  - 3p - 12bS_2^2 } \right)^2
        \right],
\end{eqnarray}
where $Z=1+\alpha\left( \rho - 3p - 12b
S_2^2\right)$. The torsion function $S_1$ in accordance with (21) is
\begin{eqnarray}\label{30}
    S_1  = -\frac{\alpha }{4Z} [\dot \rho
    - 3 \dot p + 12f_0 \omega H S_2^2
    -12( {2b - \omega f_0 } ) S_2 \dot S_2].
\end{eqnarray}
The torsion function $S_2^2$ according to (22) satisfies quadratic algebraic
equation having the following solution
\begin{eqnarray}\label{31}
 S_{2}^{2}  = \frac{\rho - 3p}{12b} + \frac
{1-(b/2f_0) (1 +  \sqrt{X})} {12b \alpha (1- \omega/4)},
\end{eqnarray}
where
\begin{equation}\label{37}
X=1+ \omega (f_0^2/b^2) [1- (b/f_0) - 2(1- \omega /4) \alpha ( \rho + 3p)]\ge
0.
\end{equation}

In order to build inflationary models we will suppose that at initial stages of
cosmological expansion HIM contain besides usual matter with energy density
$\rho_m >0$ and pressure $p_m \ge 0$ also scalar field $\phi$ with potential
$V=V(\phi)$. By using minimal coupling with gravitational field matter
components satisfy the same equations as in GR. By neglecting the interaction
between matter components, we obtain in accordance with (23) the following
equations:
\begin{equation}\label{31}
\dot{\rho}_m+3H\left(\rho_m+ p_m\right)=0,
\end{equation}
\begin{equation}\label{32}
\ddot{\phi}+3H\dot{\phi}=-\frac{\partial V}{\partial \phi}.
\end{equation}
Expressions for total energy density $\rho$ and pressure $p$ in eqs. (27)-(31)
have the form:
\begin{eqnarray}\label{33}
\rho=\frac{1}{2}\dot{\phi}^2+V+\rho_m \quad (\rho>0), & &
p=\frac{1}{2}\dot{\phi}^2-V+p_m.
\end{eqnarray}
By using (30)-(33) the torsion function $S_1$ can be expressed in the following
form
\begin{eqnarray}\label{35}
S_1  = -\frac{3f_0 \omega \alpha }{4bZ} (HD+E)  ,
 \end{eqnarray}
where
\begin{eqnarray}\label{36}
D = \frac{1}{2} \left(3\frac{d p_m}{d \rho_m}-1 \right) \left(\rho_m+p_m\right)
\nonumber\\
 +\frac{1}{3}\left(\rho_m- 3p_m\right)+\frac{2}{3}\dot{\phi}^2+\frac{4}{3} V
-\frac{b}{6f_0\alpha (1-\omega/4)} \sqrt{X}
\nonumber\\
+\frac{1-\omega (f_0/2b)}{2\sqrt{X}}\Big [\left(3\frac{d p_m}{d\rho_m}+1
\right) \left(\rho_m+p_m\right)+ 4 \dot{\phi}^2 \Big]
\nonumber\\
+ \frac{1-(b/2f_0)}{3\alpha (1-\omega/4)},
\nonumber\\
E = \left(1+ \frac{1-\omega (f_0/2b)}{\sqrt{X}} \right)\frac{\partial
V}{\partial \phi}\dot{\phi},
\nonumber\\
Z =  \frac{-\omega/4 + (b/2f_0)(1+ \sqrt{X})} {1-\omega/4}.
\end{eqnarray}
Then cosmological equation (27) leads to the following expression of the Hubble
parameter $H$:
\begin{eqnarray}\label{37}
H_{\pm}=\Bigg[-\frac{3f_0 \omega \alpha}{2b Z}E \pm \Big(\frac{1}{6f_0 Z}
       \Big [\rho  + 6(f_0 Z -b) S_2^2
       \nonumber\\
            + \frac{[1-(b/2f_0)(1+\sqrt{X})]^2} {4 \alpha (1-\omega/4)^2 }
            \Big]
            -\frac{k}{R^2}\Big)^{1/2} \Bigg]\Big(1+ \frac{3f_0 \omega \alpha}{2b Z}
D\Big)^{-1}.
\end{eqnarray}

Equations of isotropic cosmology obtained above lead to principal consequences
in behaviour of HIM at the beginning of cosmological expansion, when energy
density and pressure are extremely high. In the case of positive values of
parameters $\omega$ ($0<\omega<4$) and $\alpha$ the restriction on admissible
values of energy density and pressure follows from condition of positivity of
$X$ (31). In the case of models filled with usual matter with energy density
$\rho_m>0$ ($p_m=p_m(\rho_m)\ge 0$) without scalar fields the equality given by
(31) determines the limiting (maximum) energy density $\rho_{max}$ of order
$(\omega \alpha)^{-1}$. The state with $\rho_m = \rho_{max}$ corresponds to a
bounce and near to this state the gravitational interaction has the character
of repulsion. According to (37) the Hubble parameter with its time derivative
near a bounce are:
\begin{eqnarray}\label{38}
& & H_{\pm}=\pm \frac{2b^2}{3f_0^2 \omega \alpha} \frac{\sqrt{X} [(1/4b)(\rho_m
+p_m) - (k/R^2)]^{1/2}}{(3\frac{d p_m}{d\rho_m}+1 ) (\rho_m+p_m)},
\nonumber\\
& & \dot{H}=\frac{4b^2}{3f_0^2\omega \alpha } \frac{(1/4b)(\rho_m +p_m) -
(k/R^2)}{(3\frac{d p_m}{d\rho_m}+1 ) (\rho_m+p_m)}.
\end{eqnarray}
$H_{-}$- and $H_{+}$-solutions describe the stages of compression and expansion
correspondingly, and the transition from compression to expansion takes place
by reaching $\rho_{max}$.

In the case of models including at initial stage of expansion also scalar
fields the condition (31) determines the domain of admissible values of matter
parameters $(\rho_m, \phi, \dot{\phi})$ limited in the space of these
parameters by surface $L$ given by equality (31). The existence of this surface
provides the regularity of corresponding HIM including inflationary models.
Near surface $L$ ($X\ll 1$) the Hubble parameter according to (37) can be
expressed in the form of expansion relative to $\sqrt{X}$ \cite{28}:
\begin{eqnarray}\label{19}
H_{\pm}=H_L (1+ k_1 \sqrt{X}+ k_2 X + k_3 X^{3/2} +...),
\end{eqnarray}
where
\begin{eqnarray}\label{39}
H_L=\frac{-2 \frac{\partial V}{\partial \phi}\phi'}{(3\frac{d p_m}{d\rho_m}+1)
\left(\rho_m+p_m\right)+4 \phi'^2}.
\end{eqnarray}
and factors $k_i$ $(i=1,2,...)$ are some functions of material parameters
$(\rho_m, p, \phi, \dot{\phi})$ defined from (36). In the case of the presence
of scalar fields the bounce takes place by reaching the state with $H=0$ ($X
\neq 0$) and the value of limiting energy density in this case is different for
various solutions.

The analysis of cosmological solutions near $L$-surface (near a bounce) shows
that the set of important physical characteristics $F$ (Hubble parameter,
torsion function $S_1$, their time derivatives, curvature functions) can be
represented in the form similar to (39):
\begin{eqnarray}\label{40}
F_{\pm}= F^{(0)}+ F^{(1/2)}\sqrt {X}+ F^{(1)} X+...,
\end{eqnarray}
where expansions coefficients $F^{(0)}$, $F^{(1/2)}$, $F^{(1)}$... are some
regular functions of material parameters \cite{28}. Remarkable feature of
isotropic cosmology built in the frame of GTRC is its total regularity. All
cosmological solutions are regular not only with respect to metric with its
time derivatives and matter parameters but also with respect to torsion and
curvature \footnote{Discussed physical consequences are essentially connected
with used restrictions on indefinite parameters $a=0$ and $2f_1-f_2=0$. Without
using these restrictions the dynamics of HIM differs essentially from that
given above \cite{33} (see \cite{34}).}. In the case of HIM containing at a
bounce essentially high scalar fields we obtain inflationary cosmological
solution containing the transition stage from compression to expansion,
inflationary stage with slow rolling scalar field and post-inflationary stage
with oscillating scalar field. Similarly to inflationary HIM with only torsion
function $S_1$ investigated in \cite{29} inflationary solutions in the case of
HIM with two torsion functions can be obtained by numerical integration of
system of equations (28), (32), (33) by given initial conditions on extremum
surface $H=0$ (one assumes that equation of state $p_m=p_m(\rho_m)$ and the
form of potential $V$ are known). Particular numerical inflationary solution is
given for flat model ($k=0$) by choosing quadratic potential for scalar field
$V=m^2 {\phi}^2/2$ and $p_m=\rho_m/3$ in \cite{30}.

Physical processes at the beginning of cosmological expansion depend
essentially on value of limiting energy density (limiting temperature)
depending on values of parameters $\alpha$ and $\omega$. From physical point of
view the role of inflationary HIM in the frame of discussed regular isotropic
cosmology differs from that of standard cosmological scenario because of the
absence of the beginning of the Universe in time. However, similarly to
inflationary cosmology built in the frame of metric theory of gravity,
inflationary scenario in GTRC explains why our Universe is homogeneous and
isotropic at cosmological scale as well as it has to explain the origin of
primordial cosmological fluctuations, which are a source of  the origin of
inhomogeneous structure of the Universe and which become apparent in the cosmic
microwave background anisotropy. In connection with this it should be noted
that the building of fluctuations theory in the frame of regular inflationary
HIM discussed above is complicated, still not solved problem. Besides
complexity of gravitational equations of GTRC, the description of gravitational
fluctuations is also essentially more complicated; so the scalar gravitational
fluctuations in such models are described besides two gauge-invariant functions
of metric fluctuations also by means of  a number gauge-invariant fluctuations
functions of the torsion tensor.

Thus isotropic cosmology built in the frame of GTRC and based on cosmological
equations (27)-(28) includes two indefinite parameters $b$ and $\omega$
satisfying the conditions $1-\frac {b}{f_0}\ll1$ and $0<\omega<4$, the third
parameter $\alpha$ can be defined by using the value of effective cosmological
constant accepted by observational data \footnote{Additional restriction on
parameter $\omega$ follows from condition of positivity of $S_2^2$ determined
by (30).}. Remainder indefinite parameters in gravitational Lagrangian (1) can
be excluded by using additional physical considerations. Thus we can use
restrictions on indefinite parameters obtained in \cite{16} from analysis of
particle content of GTRC in linear approximation and exception of ghosts and
tachyons \footnote{The strict analysis of particle content has to be connected
with consideration of gravitational perturbations above the vacuum space-time
having the structure of Riemann-Cartan continuum with de Sitter metric.
However, the deviation of the structure of the vacuum space-time from Minkowski
space-time, which is essential at cosmological scale, can be unimportant by
local analysis given in \cite{10} because of smallness of values of parameter
$H$ and torsion for the vacuum.}. Restrictions on indefinite parameters
obtained in the frame of isotropic cosmology are compatible with the following
conditions: $f_1=f_2=f_3=f_4=0$ and
\begin{eqnarray}\label{17}
a_1=f_0 (1-x),\qquad a_2=2f_0 (1-x),
\nonumber \\
 a_3=-\frac{4}{3}f_0(1-x), \qquad f_5=3f_0^2 \alpha \omega,
\nonumber \\
  f_6=f_0^2 \alpha (1-\omega), \qquad (x=1- \frac{b} {f_0}).
\end{eqnarray}
The particle content of GTRC with such restrictions on indefinite parameters
includes besides massless graviton massive particles with spin-parity $2^{+}$.
In this theory the second condition (4) for parameters $f_i$ is not valid, and
the gravitational potential $\phi$ for material point of mass $M$ is determined
by (6) with the following values of $k$ and $m$: $k=\frac{x}{1-x}$,
$m=\frac{x}{3f_0 \alpha (1-x)\omega}$. Because the parameter $x$ is small we
have $k\ll1$ and hence the applying of potential (6) will give the same result
as in GR at least in the Solar system.

It should be noted that equations of discussed GTRC have a number of solutions
which are unacceptable from physical point of view. In particular, any vacuum
solution of GR with vanishing torsion is exact solution of GTRC independently
on values of indefinite parameters $f_i$ and $a_k$ \cite{16} while solutions of
GTRC far from spatially limited systems have to tend to the vacuum solution
with non-vanishing torsion. In the case of HIM there are unacceptable solutions
corresponding to the choice of the sign "minus" before $\sqrt{X}$ in expression
(10). In connection with this we have to state the criterion, which allows to
distinguish acceptable solutions from unphysical ones. Such criterion can be
based on investigation of solutions at asymptotics: far from spatially limited
systems and at asymptotics of flat cosmological models solutions of GTRC have
to tend to the vacuum solution in the form of corresponding Riemann-Cartan
continuum.

\section{CONCLUSION}

The investigation of isotropic cosmology built in the framework of GTRC shows
that this theory of gravity offers opportunities to solve some principal
problems of GR. It is achieved by virtue of the change of gravitational
interaction by certain physical conditions in the frame of GTRC in comparison
with GR. The change of gravitational interaction is provoked by more
complicated structure of physical space-time, namely by space-time torsion. In
the frame of GTRC the gravitational interaction in the case of usual
gravitating matter with positive values of energy density and pressure can be
repulsive. The effect of gravitational repulsion appears at extreme conditions
and also in situation when energy density is very small and vacuum effect of
gravitational repulsion is essential. This allows to solve the problem of
cosmological singularity and to explain accelerating cosmological expansion at
present epoch without using the notion of dark energy. The investigation of
gravitational interaction in the case of astrophysical objects is of direct
physical interest. The effect of gravitational repulsion at extreme conditions
has to prevent the collapse of massive objects and the formation of singular
black holes \cite{31}. The investigation of gravitational interaction at
astrophysical scale in the frame of GTRC is of great interest also in
connection with the dark matter problem.

\label{lastpage-01}
\end{document}